\documentclass[showpacs,amsmath,amssymb,prb,aps,twocolumn]{revtex4}
\usepackage{graphicx}
\begin{document}
\title {A new collective mode in the fractional quantum Hall liquid}
\author {I. V. Tokatly$^{1,2}$}
\affiliation{$^1$Lerhrstuhl f\"ur Theoretische Festk\"orperphysik, Universit\"at Erlangen-NÂurnberg, Staudtstrasse 7/B2, 91058 Erlangen, Germany\\
$^2$Moscow Institute of Electronic Technology, Zelenograd, 124498 Russia}
\author{G. Vignale}
\affiliation{Department of Physics
and Astronomy,  University of Missouri, Columbia, Missouri
65211, USA}
\email{vignaleg@missouri.edu}
\date{\today}  
\begin{abstract}We apply the methods of continuum mechanics to the study of the  collective modes of the fractional quantum Hall liquid.  Our main result is that at long wavelength there are {\it two} distinct modes of oscillations, while previous theories predicted only {\it one}.  The two  modes are shown to arise from the internal dynamics of shear stresses created by the Coulomb interaction in the liquid. Our prediction is supported by recent light scattering experiments, which report the observation of two long-wavelength modes in a quantum Hall liquid.
\end{abstract}
\maketitle

The two-dimensional electron liquid in semiconductor heterostructures is a remarkable many-body system.  At low temperature and high magnetic field it exhibits the fractional quantum Hall effect\cite{Tsui82,QHE} whereby the Hall resistance  assumes a universal value, independent of material parameters.   This effect, in many ways comparable to superconductivity and  Bose-Einstein condensation, is understood as the manifestation of a collective state -- the incompressible quantum Hall liquid.\cite{Laughlin83,Jain89}  The electrons in this state do not behave like independent particles, but respond   to external perturbations as a single entity: in particular, they exhibit collective density oscillations (collective modes) similar to phonons in a solid, with the crucial difference that  in the long wavelength limit the frequency tends to a finite value (the gap).\cite{Girvin86,Kamilla97,Scarola00}  Light scattering experiments\cite{Pinczuk93,Pinczuk98,Kang01} have confirmed  the existence of  a gapped collective mode, whose frequency decreases with decreasing wavelength.   However quite recently they have also revealed  the existence of a second mode,\cite{Hirjibehedin05}  whose frequency increases with decreasing wavelength (the two modes appear to merge together in the limit of infinite wavelength).  What is the origin of the second mode?  In the past there have been suggestions that two long-wavelength modes might arise from some interaction between free and bound {\it pairs} of short wavelength excitations, known as rotons.\cite{He96,Ghosh01}   Here we provide a sharper answer to the question, based on a recently developed continuum mechanics approach\cite{Tokatly05} to the dynamics of incompressible liquids.   The existence of two collective modes is shown to be a direct consequence of the  dynamics of the shear stresses created by the Coulomb interaction in the incompressible liquid. In both modes a small element of the liquid performs a  circular motion, driven by the combined action of the ordinary shear force and the Lorentz shear force (the nature of these forces will be elucidated below).  The sense of rotation is opposite for the two modes.  In the standard mode the two shear forces act in the same direction, while in the new mode they act in opposite directions.  Our approach enables us to calculate the dispersion and the splitting of the two modes in terms of a few elastic moduli, which are determined from sum rules and self-consistency conditions. The calculated dispersions are in qualitative agreement with experiment.\cite{Hirjibehedin05}

In a collective mode each small element of the liquid performs an oscillatory motion of angular frequency $\omega$ about its equilibrium position  ${\bf r}$.   We denote by ${\bf u} ({\bf r},\omega)$ (a complex vector in the $x-y$ plane) the amplitude of this motion.  It obeys an equation of motion, which follows from the conservation laws for the number of particles and the momentum:\cite{Tokatly05}
\begin{equation} \label{eom}
-m\omega^2 {\bf  u} - i\omega e B{\bf u}\times {\bf \hat z}+\frac{1}{n} \nabla \cdot \stackrel{\leftrightarrow}{\bf P} =0~,
\end{equation}
where  $m$ and $-e$ are the mass and the charge of the electron,  $B$ is the magnetic field (which points in the positive $z$ direction), ${\bf \hat z}$ is a unit vector in the $z$-direction, $n$ is the equilibrium density of electrons, $\nabla$ is the gradient operator, and $\stackrel{\leftrightarrow}{\bf P}$ is a symmetric rank-2 tensor, known as the {\it stress tensor}, whose divergence $\nabla \cdot \stackrel{\leftrightarrow}{\bf P}$ is the negative of the areal force exerted on a small element of the liquid by the liquid that surrounds it.  

As in the standard elasticity theory\cite{LandauVII}, the stress tensor is  directly proportional to the strain tensor --  a symmetric rank-2 tensor formed from the spatial derivatives of the displacement (Hooke's law).  The general form of such proportionality relation is
\begin{equation}\label{defP}
P_{ij}= -\frac{1}{2}\sum_{kl}Q_{ijkl}(\nabla_k u_l+\nabla_l u_k)~,
\end{equation}
where the indices $i,j,k,l$ denote Cartesian components $x$ or $y$,
$P_{ij}$ is the $ij$ component of $\stackrel{\leftrightarrow}{\bf P}$,
$\nabla_k$ is the spatial derivative with respect to $x$ or $y$, and
$Q_{ijkl}$ is the $ijkl$ component of the dynamic
(i.~e. frequency-dependent) {\it elasticity tensor} -- a rank-4 tensor
which contains all the information about the elastic properties of the
liquid. We will see that a particular, strongly non-instantaneous
character of the stress-strain relation (\ref{defP}) is the most
important feature of quantum Hall liquids, which makes them so
remarkable representatives of condensed matter systems.

Symmetry plays a crucial role in our analysis.  Rotational symmetry in the $x-y$ plane, combined with the presence of an axial vector ($B{\bf \hat z}$) perpendicular to that plane, dictates the following general form for the elasticity tensor:
\begin{eqnarray}\label{defQ}
Q_{ijkl}=(K+\mu)\delta_{ij}\delta_{kl}+\mu\delta_{ik}\delta_{jl}
-i\omega \Lambda \left(\epsilon_{ik}\delta_{jl}+\epsilon_{jk}\delta_{il}  \right)~,\nonumber\\
\end{eqnarray}
where $\delta_{ij} = 1$ if $i=j$ and $0$ otherwise, while $\epsilon_{ij}=0$ if $i=j$, and $\epsilon_{xy}= 1$,  $\epsilon_{yx}=-1$.

The elastic moduli  $K$, $\mu$, and $\Lambda$ are functions of frequency.  $K$ and $\mu$ are familiar from ordinary  elasticity theory:\cite{LandauVII} they are the proper bulk modulus\footnote{The definition of the bulk modulus  in the presence of a magnetic field requires extra care.    $K$ is the stiffness of an elemental area of the liquid against a compression that reduces the area but not the magnetic flux passing through it.  In other words, the magnetic flux lines are caught inside the elemental area in such a way that the local filling factor of the Landau level does not change during the compression. One way to understand the process is to observe that by keeping the magnetic flux  constant  within the area we avoid the generation of boundary currents driven by Faraday's electromotive force. These edge currents are explicitly taken into account by the Lorentz force term in Eq.~(\ref{eom}), and should not be counted twice.  These qualitative considerations can be formalized through a careful analysis of the kinetic equation in the presence of a magnetic field.} and the shear modulus respectively.    
On the other hand,  $\Lambda$ is a novel feature of the present theory. We call it  {\it magnetic shear modulus}, because  it creates a  ``Lorentz shear stress",  which is proportional to the rate of change of the shear strain.  Intuitively, the magnetic shear stress produces a force which ``squeezes together" two parallel streamlines on which the electrons move with different velocities.

Substituting Eq.~(\ref{defQ}) in Eq.~(\ref{defP}) and this in Eq.~(\ref{eom}), we obtain a linear differential equation for the displacement amplitude.  Assuming a spatial dependence of the form ${\bf u}({\bf r},\omega) \sim {\bf u} e^{i{\bf q}\cdot{\bf r}}$, where ${\bf q}$ is a two-dimensional wave vector, and going to the  limit of high magnetic field, $\omega_c \equiv \frac {eB}{m}\gg \omega$, we arrive at  the equation of motion:
\begin{equation}\label{eom.2}
- i \omega  \left(eBn + \Lambda q^2\right){\bf u}\times {\bf \hat z}+  K{\bf q} ({\bf q}\cdot {\bf u})+ \mu q^2 {\bf u}=0~.
 \end{equation}
 Non-trivial solutions of this linear equation exist only when the frequency $\omega$ has the proper $q$-dependent values:  this gives the dispersion of the collective modes.

In order to proceed we need explicit expressions for  the moduli $K$, $\mu$ and $\Lambda$ as functions of frequency.  In the long wavelength limit ($q=0$) these can be obtained from a well-known fluctuation-dissipation theory, which relates the elasticity tensor to the correlation that exists between stresses in different parts of the system -- the so-called stress-stress correlation function. This correlation function has a pole whenever the frequency matches an excitation energy of the proper symmetry -- quadrupolar in this case, since the stress tensor is a symmetric rank-2 tensor.\footnote{There is also a dipolar excitation, which is driven by the bulk Lorentz force and entails a rigid motion of the electron liquid, without internal stress.   This is known as Kohn's mode, and occurs at the cyclotron frequency $\omega_c$. Such a mode is excluded from consideration when we take the limit $\omega \ll \omega_c$.}  Our fundamental assumption is that there is only {\it one} such excitation at $
 q=0$ at a frequency $\omega = \Delta$.  Of course, this assumption is only justified in an incompressible liquid state, such as the one that sustains the fractional quantum Hall effect at special densities.  Then the theory implies that  the elastic moduli have the following form:
\begin{eqnarray}\label{Moduli}
K(\omega) &=&K\nonumber\\
\mu(\omega) &=& \mu_\infty +\frac{\mu_\infty \Delta^2}{\omega^2 - \Delta^2}= \frac{\mu_\infty \omega^2}{\omega^2 - \Delta^2}~,\nonumber\\
\Lambda(\omega)&=&\frac{\Lambda_0 \Delta^2}{\omega^2 - \Delta^2}~, 
\end{eqnarray}
where $\mu_\infty$ is the high-frequency shear modulus (i.e. the shear modulus at a frequency much larger than $\Delta$, but much smaller than $\omega_c$), and $\Lambda_0$ is the magnitude of the magnetic shear modulus at zero frequency.\footnote{In principle, the elastic moduli are also functions of the wave vector. While this dependence can affect the dispersion of the collective modes to order $q^2$ it can be shown not to affect our main qualitative results, namely the existence of two collective modes, and the magnitude of the frequency splitting to order $q^2$.}  Notice that while $\mu$ and $\Lambda$ exhibit a resonance at $\omega = \Delta$, $K$ remains finite and independent of frequency.  For this reason $K$ can be neglected in the small-$q$ limit  (see Eq.~(\ref{eom.2})), leaving the shear moduli masters of the field.\footnote{This conclusion remains valid when the infinite range of the Coulomb interaction is taken into account, which leads to $K$ diverging as $q^{-1}$
 in the long-wave length limit.}   Furthermore, it follows from the positivity of dissipation, and we will verify explicitly below, that $\mu_\infty$ and $\Lambda_0$  satisfy the inequality
\begin{equation}\label{Lambda.limit}
\Lambda_0 \Delta \leq \mu_\infty~,
\end{equation}
where the equality holds for an ideal system of non-interacting electrons.  For real electrons,  which interact via the Coulomb force, one expects  $\Lambda_0 \Delta < \mu_\infty$.  We will see that this is crucial to the occurrence of the new collective mode.

The solution of the equation (\ref{eom.2}) with the elastic moduli of
Eq.~(\ref{Moduli}) yields a two-branch excitation spectrum presented
in Fig.~1. The lower branch with a roton-like minimum corresponds to a
well known magneto-plasmon mode
\cite{Girvin86,Kamilla97,Scarola00,Pinczuk93,Pinczuk98}, while the
upper branch with a positive dispersion is a new collective excitation
predicted by the present theory, and observed experimentally in Ref.~\cite{Hirjibehedin05}. 

The origin of the two collective modes can be easily understood from a direct analysis of the
equation of motion~(\ref{eom.2}) in the limit of small $q$.  In this limit Eq.~(\ref{eom.2}) becomes
isotropic and its solutions are straightforwardly found to be 
\begin{equation}\label{eigenvectors.2}
{\bf u}_- =\frac{1}{\sqrt{2}} \left(\begin{array}{c}1\\ 
+i \end{array}\right)~,~~~~~{\bf u}_+=\frac{1}{\sqrt{2}} \left(\begin{array}{c}1\\ 
-i \end{array}\right)~.
\end{equation}
Notice that the $x$ and $y$ components of ${\bf u}$ are $90^0$ out of phase: ${\bf u}_-$ describes a counter-clockwise rotation of each element of the liquid and  ${\bf u}_+$  a clockwise rotation.

 Substituting  ${\bf u}_-$ in Eq.~(\ref{eom.2}), making use of Eq.~(\ref{Moduli}) for the elastic moduli,  and setting $\omega=\Delta$ everywhere except in the denominators we arrive at 
 \begin{equation}\label{ForceBalance1}
 \Delta e Bn {\bf u}_- = -\frac{\mu_\infty\Delta^2+\Lambda_0\Delta^3}{\omega^2 - \Delta^2} q^2 {\bf u}_-~.
 \end{equation}
 This equation states that the bulk Lorentz force acting on a liquid element (left hand side) is balanced  by the sum of the regular shear force and the Lorentz shear force exerted by the surrounding liquid (right hand side).    For perfect balance to occur the frequency must be below the gap and be given by
 
\begin{eqnarray}\label{eigenfrequency1}
\omega_{-}(q) &\simeq& \Delta- \frac{1}{2\hbar n}\left(\mu_\infty + \Lambda_0\Delta\right) (q \ell)^2~,\nonumber\\
\end{eqnarray}
where $\ell \equiv  \sqrt{\frac{\hbar}{eB}}$ is the magnetic length and $\hbar$ is the Planck constant.
Thus we see that this collective mode has negative dispersion.  This is the ``old" magneto-plasmon mode of the fractional quantum Hall liquid.\cite{Girvin86,Kamilla97,Scarola00,Pinczuk93,Pinczuk98}  Notice that the two shear forces are parallel and in phase and both push away from the center of rotation, while the bulk Lorenz force pulls towards it (see Fig.~1).

Let us now consider the other solution ${\bf u_+}$, in which the liquid elements rotate clockwise.  Repeating the same steps as above we arrive at the force balance equation
 \begin{equation}\label{ForceBalance2}
 -\Delta e Bn{\bf u}_+ = -\frac{\mu_\infty\Delta^2-\Lambda_0\Delta^3}{\omega^2 - \Delta^2} q^2{\bf u}_+~.
 \end{equation}
  Since $\mu_\infty > \Lambda_0\Delta$ we see that this condition can be satisfied only for frequencies above the gap, and the solution is
\begin{eqnarray}\label{eigenfrequency2}
\omega_{+}(q) &\simeq& \Delta+ \frac{1}{2 \hbar n}\left(\mu_\infty - \Lambda_0\Delta\right) (q \ell)^2~.
\end{eqnarray}

Thus, this second mode has a positive dispersion, with a curvature that depends on the difference
$\mu_\infty - \Lambda_0\Delta$.  Now the bulk Lorentz force and the shear Lorentz force point away from the center of the circle, while the ordinary shear force is directed towards the center (see Fig.~1). Thus, the regular shear force and the Lorentz shear force act in opposite directions.  For this reason the dispersion is weaker than in the - mode.
The difference between the frequencies of the two modes is  
\begin{equation}\label{frequencydifference}
\omega_+(q)-\omega_-(q) =\frac{\mu_\infty}{\hbar n} (q\ell)^2~,
\end{equation}
which depends {\it only} on the high-frequency shear modulus, {\it not} on $\Lambda_0$ and $K$.   

{\bf In Fig. 1 we plot the dispersions of the collective modes obtained from the solution of  Eq.~(\ref{eom.2}) with the elastic moduli given by Eq.~(\ref{Moduli}).  The determination of  $\mu_\infty$ and $\Delta$ will be discussed below.  $\Lambda_0$ and  $K$ are fixed by requiring that the dispersion of the - mode has a magnetoroton minimum at the correct values of $q\simeq1.4 \ell^{-1}$ and $\omega\simeq\Delta/2$.\cite{Pinczuk98,Kang01}   Notice that we are plotting these dispersions not only for small $q$, where Eqs.~(\ref{eigenfrequency1}) and~(\ref{eigenfrequency2}) are valid, but for {\it all} $q$.  Although the effective elasticity theory is on firm ground only for small $q$, it is rewarding to see that the dispersion of the $-$ mode is sensible for all $q$.  On the other hand, we believe that the  $+$ mode exists only for very small $q$, after which it merges into a continuum of more complex excitations, such as two-roton excitations.}

From the solution of the equation of motion~(\ref{eom.2}) (augmented by an external forcing term on the right hand side) we can easily derive the long-wavelength behavior of the density-density response function, the quantity that is most directly probed by inelastic light scattering experiments:\cite{Pinczuk93,Pinczuk98,Hirjibehedin05} 
 \begin{equation}\label{chinn}
\chi(q,\omega)=\frac{(q\ell)^4}{2\hbar^2}\left\{\frac{\mu_\infty -\Lambda_0 \Delta}{\omega^2-\omega_+(q)^2}+\frac{\mu_\infty +\Lambda_0 \Delta}{\omega^2-\omega_-(q)^2}\right\}~.
\end{equation}
Notice that the static response function $\chi(q,0)$ vanishes as $q^4$ for small $q$, proving that the system is incompressible even though the bulk modulus is finite and does not even enter the long-wavelength dispersions.  The dynamical response function has two resonances at $\omega = \omega_-(q)$ and $\omega=\omega_+(q)$, but notice that the two resonances have different strengths, proportional to    $\mu_\infty \pm\Lambda_0 \Delta$ respectively.  
Positivity of dissipation requires that the strength of a resonance (also known as oscillator strength) be positive:  hence the inequality~(\ref{Lambda.limit}) is confirmed.

It should be noted that the $+$ mode becomes weaker as $\Lambda_0 \Delta$ approaches $\mu_\infty$, and disappears when the equality $\Lambda_0\Delta=\mu_\infty$ attains.   In a gas of non-interacting electrons/composite fermions  in the lowest Landau level one can easily show that  indeed $\Lambda_0\Delta=\mu_\infty$.  The reason for such a finely tuned relation between $\Lambda$ and $\mu$ is that they both arise from the {\it kinetic} stress tensor operator, $P_{ij} \propto v_iv_j$ where ${\bf v}$ is the velocity operator. In the lowest Landau level, this satisfies the constraint $P_{xx}-i P_{xy} = 0$ because the clockwise component of the velocity operator ($v_x - iv_y$) annihilates every state in the lowest Landau level.  This fine tuning is destroyed as soon as the interaction part of the stress tensor is included.  Therefore,  the very possibility of observing the +  mode depends on the Coulomb interaction between the electrons (or composite fermions): without it Eq.~\ref{ForceBalance2} would not have a solution.

\begin{figure}\label{Dispersion}
\includegraphics[width=8cm]{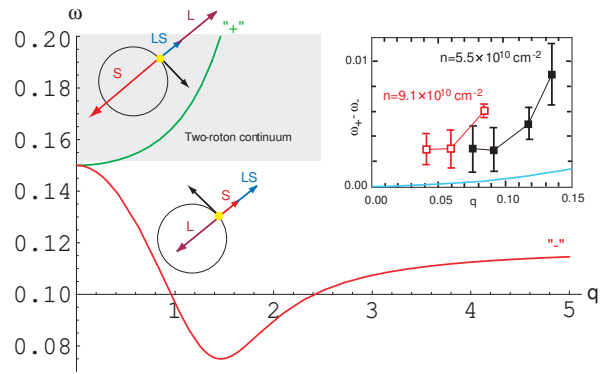}
\caption{{\bf Dispersion of the collective modes calculated from Eqs.~(\ref{eom.2}) and~(\ref{Moduli}) with $\nu=1/3$, $\Delta=0.15$,  $\mu_{\infty}/n=0.075$,  $\Lambda_0 /(\Delta\mu_\infty) =0.7$ and $K/n=0.025$.  The $+$ and  $-$  modes are shown in green and red respectively. $q$ is in units of $\ell^{-1}$ and $\omega$ is in units of $\frac{e^2}{\hbar \epsilon_b \ell}$. Next to each dispersion we show schematically the direction of rotation of the displacement field and the balance of the forces that sustain the mode:  ``L" denotes the Lorentz force, ``S"  the ordinary shear force, and  ``LS" the Lorentz shear force. The grey region denotes the conjectured two-roton continuum, which limits the dispersion of the $+$ mode.  The inset compares the calculated frequency splitting (blue line) with the experimental data of Ref.~\cite{Hirjibehedin05} (solid and open squares) for two samples of different electron density.  The vertical bars indicate the experimental uncertainty.}}
\end{figure}    

We can calculate the high-frequency shear modulus $\bar \mu_\infty$, and hence the frequency splitting,  in the following manner.  Integrating the imaginary part of $\chi$ over frequency yields the static structure factor 
\begin{equation}
S(q) =  \frac{\mu_\infty}{2 n \hbar \Delta}  (q \ell)^4~.
\end{equation}
Comparing this with the structure factor computed directly from the Laughlin ground-state wave function\cite{Laughlin83} ($S(q) = \frac{1-\nu}{8 \nu} (q\ell)^4$  when the filling factor $\nu \equiv 2 \pi n \ell^2$ is the inverse of an odd integer)    we arrive at the following identification:
\begin{equation}
\frac{\mu_\infty}{\hbar n} = \Delta \frac{1-\nu}{4 \nu}~,
\end{equation}
and thus
\begin{equation}\label{frequencydifference}
\omega_+(q)-\omega_-(q) =  \Delta \frac{1-\nu}{4 \nu} (q\ell)^2~.
\end{equation}
So at $q=0.1 \ell^{-1}$ and $\nu=1/3$ we predict a splitting $200$ times smaller than the $q=0$ gap.    The value of  $\Delta$ can be determined self-consistently from the requirement that the response function~(\ref{chinn}) satisfy the f-sum rule in the lowest Landau level\cite{Girvin86}: $-\pi^{-1}\int_{0}^\infty \omega \Im m \chi(q,\omega) d \omega =\bar f(q)$, where $\bar f(q)$ is a known functional of the static structure factor, whose explicit form is given in Ref.~\cite{Girvin86}.   This requirement  gives  $\hbar \Delta=0.15~e^2/\epsilon_b\ell$, from which  we get a splitting of  $\hbar \Delta (q\ell)^2/2 =0.00075 ~e^2/\epsilon_b\ell$ at $q=0.1 \ell^{-1}$.    This is smaller than what is seen in the experiment of Ref.~\cite{Hirjibehedin05},  but still in qualitative agreement with it. {\bf  The agreement with experiment might be improved by including more realistic features, such as the contribution of high energy excitation to the sum rules, the mixing of higher Landau levels, and the finite width of the samples. Finally one cannot exclude the possibility that the splitting persists at $q=0$, i.e. that the stress tensor itself has two resonance of opposite chirality at $q=0$.} 

In conclusion, we have presented an effective elasticity theory that explains the occurrence of {\it two} collective modes in the fractional quantum Hall liquid.  The essential improvement upon the  theory of Ref.~\cite{Girvin86} is the recognition that the single-mode approximation should be done on the stress tensor, rather than on the density-density response function.  This is how we get two modes instead of one.  The strength of the theory lies in the ease with which it allows one to calculate analytically the  dispersion of the collective modes.  With a proper choice of the elastic moduli one obtains excellent results not only at long wavelengths but also at large and intermediate wavelengths (see Fig. 1). 

{\it Acknowledgements.} This work was supported by NSF Grant No.
DMR-0313681. We thank the authors of Ref.~\cite{Hirjibehedin05} for kindly making their data available to us, and Prof. Jainendra K. Jain for many discussions and invaluable advise.


\begin{thebibliography}{25}
\bibitem{Tsui82}D. C. Tsui, H. L.  St\"ormer, and  A. C. Gossard, ~Phys. Rev. Lett. ~{\bf 48}, 1559 (1982).
\bibitem{QHE} R. E. Prange and S. M. Girvin, editors~{\it The Quantum Hall Effect} (Springer-Verlag, New York, 1987); S. Das Sarma and  A. Pinczuk, editors ~{\it Perspectives in Quantum Hall Effect} (Wiley, New York, 1990).
\bibitem{Laughlin83}R. B. Laughlin,~Phys. Rev. Lett. ~{\bf 50}, 1395 (1983).
\bibitem{Jain89}J. K. Jain, Phys. Rev. Lett. ~{\bf 63}, 199 (1989).
\bibitem{Girvin86} S. M. Girvin, A. H. MacDonald, and P. M. Platzman,~Phys. Rev. B~{\bf 33}, 2481 (1986).
\bibitem{Kamilla97} R. K. Kamilla and and J. K. Jain, ~Int. J. Mod. Phys. B~{\bf 11}, 2621 (1997).
\bibitem{Scarola00} V. W. Scarola,  K. Park, K., and J. K Jain, ~Phys. Rev. B~{\bf 61}, 13064 (2000).
\bibitem{Pinczuk93}A. Pinczuk,  B. S. Dennis, L. N. Pfeiffer, and K. W. West,~Phys. Rev. Lett.~{\bf 70}, 3983  (1993). 
\bibitem{Pinczuk98}A. Pinczuk, B. S. Dennis,  L. N. Pfeiffer,  and K. W. West,~Physica B~{\bf 249}, 40 (1998). 
\bibitem{Kang01}M. Kang,  A. Pinczuk,  B. S. Dennis,  L. N. Pfeiffer,  and K. W. West,~Phys. Rev. Lett.  {\bf 86}, 2637 (2001).
\bibitem{Hirjibehedin05} C. F. Hirjibehedin,  I.  Dujovne,  A. Pinczuk,  B. S. Dennis,  L. N. Pfeiffer,  and  K. W. West,~Phys. Rev. Lett. {\bf 95}, 066803 (2005).
\bibitem{Tokatly05} I. V. Tokatly, cond-mat/0512706;~Phys. Rev. B~{\bf 73}, 205340 (2006).
\bibitem{LandauVII}L. D. Landau,   and  E. Lifshitz,  {\it Theory of Elasticity},  {\it Course of Theoretical Physics}, Vol. 7, 3rd ed.   (Pergamon Press, Oxford, 1986).
\bibitem{He96}S. He,  and P. M. Platzman,~P. M., Surf. Sci. {\bf 361/362}, 87 (1996).
\bibitem{Ghosh01} T. K. Ghosh and G. Baskaran,~Phys. Rev. Lett. {\bf 87},186803 (2001).
\end{thebibliography}
\end{document}